# Switching Economics for Physics and the Carbon Price Inflation: Problems in Integrated Assessment Models and their Implications

## Sgouris Sgouridis[1], Abdulla Kaya, Denes Csala

*Masdar Institute of Science and Technology, Abu Dhabi, United Arab Emirates*

## Abstract

Integrated Assessment Models (IAMs) are mainstay tools for assessing the long-term interactions between climate and the economy and for deriving optimal policy responses in the form of carbon prices. IAMs have been criticized for controversial discount rate assumptions, arbitrary climate damage functions, and the inadequate handling of potentially catastrophic climate outcomes. We review these external shortcomings for prominent IAMs before turning our focus on an internal modeling fallacy: the widespread misapplication of the Constant Elasticity of Substitution (CES) function for the technology transitions modeled by IAMs. Applying CES, an economic modeling approach, on technical factor inputs over long periods where an entire factor (the greenhouse gas emitting fossil fuel inputs) must be substituted creates artifacts that fail to match the S-curve patterns observed historically. A policy critical result, the monotonically increasing cost of carbon, a universal feature of IAMs, is called into question by showing that it is unrealistic as it is an artifact of the modeling approach and not representative of the technical substitutability potential nor of the expected cost of the technologies. We demonstrate this first through a simple but representative example of CES application on the energy system and with a sectoral discussion of the actual fossil substitution costs. We propose a methodological modification using dynamically varying elasticity of substitution as a plausible alternative to model the energy transition in line with the historical observations and technical realities within the existing modeling systems. Nevertheless, a fundamentally different approach based on physical energy principles would be more appropriate.

---

[1] Corresponding Author: Associate Professor, Masdar Institute of Science and Technology, P.O. Box 54224, Abu Dhabi, United Arab Emirates, Tel: +971 2 810 9156, Fax: +971 2 810 9901, email: ssgouridis@masdar.ac.ae

# 1    Introduction

Integrated assessment models (IAM) are climate economy models that are widely used for tying the evolution of the climate system to the economic system. They are state-of-the-art attempts to map the dynamic effects of climate change on the economy and vice versa, projecting them several decades into the future. Doing so allows the comparison of the costs of climate change against the costs of preventative actions, thus providing information to decision makers to inform policy development. One consistent output of the models is that in order to combat climate change, carbon prices should continue to increase throughout this century reaching to very high levels towards the end. We investigate the underlying economic model that leads to this result and discuss why the commonly applied formulation constant elasticity of substitution (CES) is unrealistic in the context of long-term technical substitution processes theoretically and practically as it does not account for the physical and technical replacement potential of alternative technologies. This paper aims first to provide a comprehensive review of the current critiques towards IAMs along several fronts before focusing on exploring the problems and resulting artifacts from using CES in technology substitutions an area that has not been previously discussed in detail.

As a starting point, Table 1 presents an indicative collection of IAMs from different years and the identifying equations for two critical aspects: the economic engine, how they model economic output, and the climate damage function, how they relate climate change to economic disruption. While there is a large proliferation of models, and Table 1 does not include several, there are fundamental similarities in how the models approach these two aspects as can be seen by the similarities in the formulations.

*Table 1 IAM models and functional formulations used for economic growth and climate damages (detailed notation explanation available at the sources)*

| Model (Reference) | Economic Engine[2] | Climate Damage Function |
| --- | --- | --- |
| **DICE(Nordhaus 1992), (Nordhaus** | Cobb-Douglas/CES | Quadratic function of temperature $\Omega(t) = \psi_1 T_{AT}(t) + \psi_2[T_{AT}(t)]^2$ |

---

[2] Please see the variable definitions of each function in the respective references, included in the bibliography with reference to the equation numbers

| | | |
|---|---|---|
| **and Sztorc 2013) pg. 10** | $Q(t) = [1 - \Lambda(t)]A(t)K(t)^\gamma L(t)^{1-\gamma}/[1 + \Omega(t)]$ | |
| **ENTICE (Popp 2004) Eq.2** | DICE with technological change<br>$Q_t = A_t K_t^\gamma L_t^{1-\gamma-\beta} E_t^\beta - p_{F,t}F_t$ | DICE |
| **ENTICE-BR (Popp 2006) Eq. 4** | ENTICE with backstop technology<br>$Q_t = A_t K_t^\gamma L_t^{1-\gamma-\beta} E_t^\beta - p_{F,t}F_t - p_{B,t}B_t$ | DICE |
| **WITCH (Bosetti et al.) Eq.1** | $Q(n,t)$<br>$= TFP(n,t)\Big[\alpha(n)$<br>$\cdot \left(K_C^{1-\beta(n)}(n,t)L^{\beta(n)}(n,t)\right)^\rho + (1$<br>$- \alpha(n)) \cdot ES(n,t)^\rho\Big]^{1/\rho}$ | DICE |
| **MARKAL (Loulou et al. 2004)Eq. 1.3-1 (Loulou and Labriet 2007)** | $X_s = A_0(B_K \cdot K_s^\rho + B_L \cdot L_s^\rho + B_E \cdot E_s^\rho)^{1/\rho}$ | $DAM(EM) = \alpha \cdot EM^\beta$ |
| **MERGE (Manne et al. 1995) Eq. 5, (Manne and Richels 2005)** | $Y = \left[a \cdot (K^\alpha L^{1-\alpha})^\gamma + b \cdot \left(E^\beta N^{1-\beta}\right)^\gamma\right]^{1/\gamma}$ | $D_{t,reg} = \alpha_{reg} \cdot \Delta T_{t,reg}^{\beta,reg} \cdot GDP_{t,reg}$<br>$W_{t,reg} = \dfrac{t_{reg} \cdot \Delta T_{t,reg}^{\delta,reg}}{1 + 100\exp\left(-\dfrac{0.23GDP_{t,reg}}{POP_{t,reg}}\right)}$ |
| **MIND Eq.3 (Edenhofer et al. 2005)** | MARKAL | no direct climate effect,<br>only based on fossil fuel scarcity |
| **MIT EPPA (Paltsev et al. 2005)** | Sectoral nested CES | Not explicit |
| **FUND (Tol 2002)** | Sectoral CES | Sectoral effects, e.g for agriculture:<br>$a_{t,r}^r = a\left(\dfrac{\Delta T_t}{0.04}\right)^\beta + \rho a_{t-1,r}^r$ |
| **REMIND-R Eq. 2 (Leimbach et al. 2009)** | MERGE & WITCH with detailed energy trade<br>$B^i(r) = \sum_t \sum_j \left(p_j^i(t)\right.$<br>$\left. \cdot \left[X_j^i(t,r) - M_j^i(t,r)\right]\right)$ | Not explicit |
| **PAGE (Hope 2006)** | Not explicit | $GDP_{i,r} = GDP_{i-1,r} \cdot \left(1 + \dfrac{GRW_{i,r}}{100}\right)^{Y_i-Y_{i-1}}$<br>$DD = \sum_{i,r} (AD_{i,r})$<br>$\cdot \prod_{k=1}^{i}\Big(1 + dr_{k,r}$<br>$\cdot \dfrac{ric}{100}\Big)^{-(Y_k-Y_{k-1})}$ |

| (Cai et al. 2015) | Nested CES. CRESH function with backstop $$\sum_i \left(\frac{Q_i}{X}\right)^{d_i} \cdot \frac{D_i}{d_i} = k$$ | DICE |
|---|---|---|
| (Baker and Shittu 2008) | $Y = [(e_c^\rho + e_{nc}^\rho)^{\frac{\rho}{\rho_e}} + H^{\rho_e}]^{1/\rho_e}$ | DICE |

In Section 2, we present the general characteristics of IAMs and the primary criticisms. Section 3 analyzes in-depth the CES function through a detailed example in order to demonstrate the method's unsuitability for applications that model technological transitions. Section 4 shows that ignoring this effect, the current state-of-the art IAMs indeed rely on this formulation leading to incongruent estimates of the marginal carbon price that is consistent among all examples. Section 5 discusses the implications of this flaw for policy analysis and suggests alternatives to this approach based on empirical research on energy transitions including a modified application of CES. Section 6 concludes recommending the use of physical as opposed to economic models for studying the long-term scenarios for climate change mitigation and energy availability.

## 2    IAM Structure and Criticisms

In order to fulfill their stated objective, IAMs must: (i) represent the economic system over decades, (ii) translate economic activity into greenhouse-gas (GHG) emissions, (iii) estimate the impact of GHG emissions on the climate usually summarized in the degrees of change of average temperature from the pre-industrial baseline, (iv) calculate how climate change impacts economic activity. Comparing the net present value of the climate damages in a baseline, business-as-usual (BAU) case against the reduction in damages but increased costs of taking mitigating actions provides an indication on the level of action that is preferable. The results are usually summarized in the form of the carbon price; a tariff that if levied on GHG emissions would provide sufficient incentives for adjustments to be made in an economically neutral way. This should not be confused with the social cost of carbon (SCC) which is the cost of GHG emissions derived through the climate damage function (step iv) represented in their current net present value. A confusion that is responsible for its severe underestimate (Ackerman and Stanton 2012). In other words the recommended carbon price should force only economically efficient adjustments to be made so that their current cost does not exceed the future costs from climate damages.

The prevailing economic view of climate change mitigation revolves around the idea that countries should deploy measures with a marginal cost that is equal to the global marginal benefit from avoiding climate change. Individual countries though have the incentive to simply reach their own marginal benefits (Aldy et al. 2003). Such a view assumes that countries can have clearly separable impacts – given that the level of interconnection of both natural and social systems is vast, this is not a tenable position. Immigration (see the ongoing 2015 immigration crisis), loss of trade (in terms of losing export and import markets because the trading partners are severely impacted), ecosystem collapse leading to cross-border pollution (see Indonesian forest fires) are just some indicative examples of such interconnections. The presumption that countries can estimate individual costs is therefore unsupportable.

At a fundamental level, IAMs focus on economic output with the implicit assumption that it is equivalent to social welfare. Even after recognizing non-monetized impacts, e.g. (Nordhaus and Sztorc 2013), they address it by adjusting the monetized impacts. By adopting GDP as the fundamental measure of the economy they inherit its limitations leading to counterintuitive results. Since GDP only traces the monetary value of consumption and investment and is agnostic to their true social value on the basis that the two should coincide. When they do not, then distortions are inevitable. For example, remediating a disaster adds to the GDP, the infamous broken window effect. Of more direct relevance, in GDP terms a reduction in the consumption of fossil fuels and mineral stocks is counted as a negative, increases in agricultural output that depletes soil fertility even if some rot in pantry shelves are counted as positive, and an investment of a given magnitude is an equal positive no matter if it is used to build an RE plant (a facility collecting a *renewable* resource) or a gas-well (a facility extracting a *depleting* resource). There are several measures proposed as alternatives but the ones that would register a difference would need to account for natural capital depletion and the associated environmental externalities.

Even after accepting that GDP is a sufficient measure, if only because policy makers - the ultimate audience of the models' results – desire it, existing IAMs generate a very wide range for the marginal price of carbon (illustrated in Fig. 4). This variation largely relates to the sensitivity of the fundamental assumptions driving the valuation process itself.

(Pindyck 2013) focuses on three of these: the choice of the discount rate, the formulation of the damage function, and the assessment of the probability of catastrophic events.

The choice of the discount rate for events far into the future plays a critical role and even identical models with a fraction of a percentage point of variation will give vastly different results. The theoretical foundation of discounting is based on individual time preferences but in climate change settings it is a question of intergenerational distribution where it can be argued that society itself has continuity not bounded by time and therefore would not have a time preference (Howarth 1998). As a result, for the distant future only very low constant discounting rates make sense (Gollier 2002). Variable discounting that allows the distant discount rate to drop close to zero while being higher for near-term accounting, like the hyperbolic discounting function, may be preferable (Karp 2005). Fundamentally, the discount rate choice should be a decision stemming from a political process on how to valuate intergenerational welfare. While the choice of discount rate is a clear-cut case of political bias inserted by the modeler in the assumptions, the two other areas also demonstrate unstated biases from the side of the modelers.

To begin with, in most IAMs, economic output is assumed to be growing at an exogenous background rate. For example the (Paltsev et al. 2005) CGE formulation results in a consistent global GDP growth that starts from around 4% in 2005 and is reduced to 2% per year by 2100. Interestingly, in an attempt to become more inclusive, a Delphi survey of experts gave the average result of 2.3% global per capita growth for 2010-2100 (Gillingham et al. 2015). In practical terms these rates mean that the average Chinese in 2100 would be able to consume three times more than the average American in 2000 in *real* terms (cf Table 19 in (Paltsev et al. 2005) and Fig. S3 in (Kriegler et al. 2015) for illustration) while the 2100 American would be consuming nine times more than her 2000 ancestor. Attaining such consumption levels let alone sustaining them in a finite world is highly unlikely in physical terms considering that the average American in 2002 had an operational ecological footprint that was already four times greater than the global per capital available biocapacity of all Earth's ecosystem's (Kitzes et al. 2008). While the financial system and mainstream economics are founded on the premise that an economy is a closed system with infinite capacity for growth the real economy is an open thermodynamic system that relies on environmental inputs for its sustenance (Sgouridis 2014). As such, when faced with resource and pollution limits the economic system will inevitably contract[3] (Ayres 1996), (Bardi 2011). The ad hoc presumption of future economic growth may in paper relieve the potential economic catastrophe of climate impacts but it is no more than wishful thinking[4].

Within this frame of practically predestined economic growth, the climate damage function (CDF) is intended to estimate the impact of climate change on economic output. To start with, climate damage functions depend on climate sensitivity – a measure that ties GHG concentrations and temperature change. The upper bounds of climate sensitivity are difficult to deduce from historical data and a fat-tail climate response, where the likelihood of high temperature changes remains substantial even for moderate GHG concentrations, cannot be precluded (Allen and Frame 2007). Now climate damage functions tie these quite uncertain estimates of climate sensitivity to how the economic system would respond. CDFs are usually presented as a percentage of the GDP that is "lost" to climate impacts but in most cases this loss does not impact the GDP background growth rate. In addition, they are estimated on speculative assumptions and tend to err on the side of low impacts. As a case in point, the climate damage function in the early DICE (see Table 1) is a quadratic function that even at 8C change will only reduce the economic output by less than 10%. These levels of temperature difference are much higher than the difference between average glacial and interglacial temperatures. When even 4C change is described as incompatible with organized society (Anderson and Bows 2010) and imply conditions beyond breaking point (Jorge Mario Bergoglio 2015) where the descriptor "hell on earth" would not be inappropriate (McIntosh 2012), yet the damage functions used presume that the economy would still hum along at 90% of the level it would have had without climate change. Notably, in its latest iteration, DICE-2013R (Nordhaus and Sztorc 2013) reverted to a quadratic damage function "fitted"[5] on a review by Tol for the range of 0-3C warming (Tol 2009). The accompanying disclaimer acknowledges the limitation that since any climate data we have are limited to the lower warming scale, the higher warming estimates are speculative. This disclaimer does not prevent the modelers from examining significantly higher warming scenarios later on. Weitzman notes that the insistence on the quadratic function is likely induced by economists' familiarity with it. More importantly though, he shows that the inherent disregard of climate damage functions to the fat-tail distribution properties of catastrophic climate change with essentially unlimited downside exposure is inappropriate and misleading (Weitzman 2009).

---

[3] Interestingly, Paltsev et al. argue that their GDP growth rate calibration is in line with the period 1950-1973 when energy and resources were for all intents and purposes unconstrained and therefore had very little effect on growth although resources impact GDP only when constrained (Ayres and Warr 2005), (Stern 2011).

[4] We were tempted to use "magical" as the adjective here. An example of the results from this attitude is the quote by Prof. Lomborg: *"…if Bangladesh was as rich as Holland, they'd be much more able to deal with the effects of [sea level] rise."* (Chivers 2013)

[5] We use quotes as in reality the number and shapes of curves that could be fitted to these estimates is practically infinite.

Tol's (Tol 2009) argument that the damage estimates are reliable since they were generated by different methods: enumerative sectoral disaggregation using physical models, statistical observations using historical country data, and Delphi surveys without substantial divergence does not address Weitzman's critique. Since the estimates refer only to a limited temperature change range they are effectively useless for assessing potentially catastrophic outcomes. More importantly, the detailed "enumerative" damage function, e.g. in the FUND model (Tol 2002) do not necessarily offer a more precise view of impacts. Firstly there are methodological problems, like functions with highly sensitive parametric behavior due to division with values close to zero (Ackerman and Munitz 2012). More importantly, providing a detailed breakdown gives a false sense of confidence that hides a few defining assumptions. First, that interactions and feedbacks between sectors are ignored – if for example, Bangladeshi agricultural land is inundated by sea level rise, agricultural output would only by impacted by temperature alone unaffected by sea level rise. Second, the implications of catastrophic events, like drastic crop failures from droughts or floods or fire, hurricanes, a monsoon failure along with the loss of glacial melt in India are not accounted for. Third, the level of impacts rely on extrapolation of moderate data points and past trends with the ultimate effect of having curves with preset growth factors similar to the economic output ones. For example change in agriculture production is based on a quadratic equation (see Table 1 2nd row) that incorporates a preset growth factor (alpha) while higher temperatures are actually increasing that "predestined" growth and is not in line with current research (Ackerman and Munitz 2012). In the longer term, this assumption is based on the availability of fertile land in the landmasses of the northern hemisphere which would be extremely costly to develop agriculturally resource-wise due to the need for fertilization and infrastructure development. Fourth, the issues of monetizing effects like the loss of ecosystem functions, in which case the cost of losing an ecosystem can be counterbalanced by everyone in the OECD skipping a $30 dollar meal, or the costs of human life are present but not discussed. Finally, the game of positive and negative effects in the end obscures the sectoral and distributional effects. As an example, the savings from the reduction in heating costs are estimated to be more or less equal to the costs of the loss of land from sea level rise. Aside from the absurdity of this equation, the fact remains that those affected by the loss of land are not the same as those who gain from having to heat their homes less.

Such unequal distribution of "benefits" and losses is further underlined by another implication of the cost-benefit analysis (CBA) approach employed in the estimates of climate-damage functions. Since a large part of damages are related to human health and mortality, the assumed value of human life influences these results. CBA ties these valuations to economic output with the ethically disturbing result of valuing lives in developed countries as much as a factor of 15 higher than a life in the developing countries (van den Bergh 2004).

The problems identified so far highlight a number of very significant shortcomings that by themselves should caution the use of IAMs as a basis for climate and energy policy. Nevertheless, these issues are related to exogenous factors, external inputs and assumptions that could, in theory, be improved to better reflect reality. The next section focuses on a potentially insidious problem as it lies at the heart of the modeling method itself and would not be fixed by a better choice of the discount rate and more accurate estimates of the climate damage function.

## 3    The Economic Allocation of Resources in IAMs: Elasticity of Substitution and Production Functions

The theoretical underpinnings and the assumptions needed for using a general equilibrium approach, e.g. market participants act independently or that markets clear instantaneously, have been questioned with regard to their applicability (c.f. (Kirman 1989), (Varoufakis et al. 2012)). Nevertheless, the problem we discuss, applying constant elasticity of substitution (CES) functions and its variants for modeling long-term technological transitions, can be present even in models that do not use a general equilibrium approach. In some instances, even technological transition models that do not use CES emulate the prototypical CES-derived behavior, Such misrepresentation of the technological process that is fundamental to any energy transition diminishes the models' policy relevance as it leads to a systematic, yet unrealistic, overestimation of the costs imposed by a transition to a zero-carbon energy system. We provide a simple implementation and then demonstrate how the same modeling artifacts are present in the results of mainstream models used for energy policy.

### 3.1    The CES Functions and its use for Technological Substitution

CES functions were introduced as a general production function form for capital and labor substitution (Arrow et al. 1961) that could provide flexibility increasing the options between a Leontief function (where the elasticity of substitution is zero) and a Cobb-Douglas function (where the elasticity of substitution is unitary). In the form shown in Eq. 1, ($\alpha$) is the share of factor of production (F) and $\rho$ defines the elasticity of substitution ($\sigma$) as $\sigma = 1/(1 + \rho)$. While the two-input form can be easily expanded into multi parameter general form, there are limitations in solving them especially if the elasticities are not equal so in practice CGE models use nested CES functions to model multiple factors of production with different elasticities of substitution (cf. Fig 3 for examples).

$$Y = [\alpha \cdot F^{-\rho} + (1 - \alpha) \cdot R^{-\rho}]^{-1/\rho} \qquad (1)$$

A CES function describes the amount of change required in one input in response to a certain amount of change in other input(s) to maintain a given utility level. An initial, perhaps trivial, observation is that CES functions are limited by the original structure of the inputs of the production function. If an input is not initially utilized and therefore not designed-in, then they obviously will not be used by the model without an intervention. I.e. if no RE is used at the time of model specification, then its share will remain zero.

A more important characteristic, is that in a CES function, the marginal rate of technical substitution (MRTS) of one factor with respect to another (i.e. $\Delta R/\Delta F$) depends on the factor proportional value (R/F) as shown in Eq.2 (derivation in the appendix).

$$\text{MRTS}_{R,\,F} = -\frac{\Delta R}{\Delta F} = -\frac{\alpha}{1-\alpha} \cdot \left(\frac{R}{F}\right)^{1+\rho} \qquad (2)$$

While the initial substitution requirements are less than unity, for a large initial share ($\alpha$) and a positive $\rho$, by its mathematical definition, CES requires exponentially increasing amounts of input from one factor to replace the other in order to maintain a constant output. **CES functions are, by design, share-preserving.** The greater the original share ($\alpha_i$) of a factor other factors would need to be utilized exponentially more to replace a marginal reduction of i while maintaining the output (Y) constant as the substitution progresses.

Initially the CES function was intended for modeling the economic output of sectors using two-factor inputs labor and capital. In this setting, neither labor nor capital would be reasonably expected to completely dominate and where the range of possible substitution between them would be limited. Bounded this way, the use of CES was perhaps justifiable and fairly representative. Nevertheless, mathematical convenience and ubiquitous use in general or partial equilibrium models of what became a conventional and commonplace approach have led to extending the use of CES functions to model processes with technical factor inputs. This extension though fails empirically because technical processes can fully substitute each other and in fact the substitution tends to become easier as the penetration of an alternative technology increases, thus changing the elasticity of the substitution. Technological substitution often takes an S-curve form that has been historically observed across economic sectors and readily identifiable in the energy domain (cf. Section 5). Had economists tried to model the transition from steam to diesel locomotives using a CES function, they would find, counterfactually that the world would still be using a large share of steam locomotives. Just substitute the factor input names in the example presented in the following section.

## 3.2 *A detailed example of a CES application for technological transition*

This example demonstrates in detail how a typical CES operates emulating the way that is applied in IAMs using computable general or partial equilibrium methods and explains why some policy-critical results are artifacts of the equation and not dictated by any real-world rationale. It is based on calculating the Eq. 1 values for modeling the substitution of a portfolio of fossil sources (F) by a portfolio of renewable energy (R) sources in order to reduce overall system emissions. These technical options are used to generate the energy (Y) required for the economic processes of a given region. The demonstration is designed to highlight the sensitivity of the function to the initial shares and the choice of the elasticity of substitution parameter. To do so, we initialize and solve the CES function assuming the cost of the fossil resources ($c_F$) as constant acting as the base unit of comparison. We calculate the amount of R needed to substitute F at any point and the corresponding price of fossil resources ($p_F$) that needs to be levied externally in order to force the transition using Eq. 2. This result of the Lagrangian function solution of the CES with the assumption of a perfectly competitive two-goods market for energy and unrestricted income is derived in the Appendix. In research practice, the preferred metric for measuring the forced replacement of F is the carbon price (not to be confused with the social cost of carbon) implemented either as a carbon tax or as a result of a cap and trade system. In this model, the difference between the fossil fuel price and marginal cost provides the level of the carbon price or tax (C) necessary to achieve this substitution (Eq. 3).

$$\frac{P_F}{P_R} = \frac{\alpha}{1-\alpha} \cdot \left(\frac{F}{R}\right)^{-\rho-1} \tag{2}$$

$$C = p_F - c_F = c_R \cdot \frac{\alpha}{1-\alpha} \cdot \left(\frac{F}{R}\right)^{-\rho-1} - c_F \tag{3}$$

Fig. 1 graphs the values of the input resources and resulting carbon tax for different initial shares (α) and elasticity of substitution (σ) values in line with the range of values used in IAMs. We make four key observations: (i) as anticipated, the substitution curve F-R shows an exponentially increasing input from R to substitute for F, (ii) the substitution rate saturates at different levels depending on the share (α) but very far from a complete substitution, (iii) the carbon price level is small in the beginning and increases exponentially as the substitution saturates, and (iv) the carbon price is *very* sensitive to the choice of the elasticity of substitution parameter.

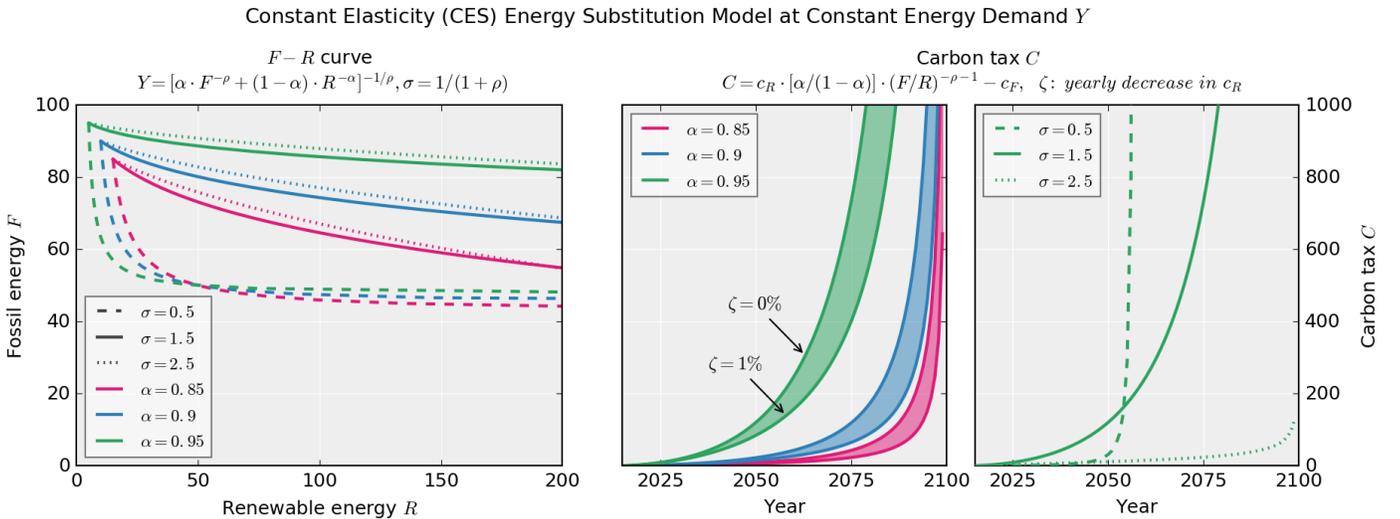

**Figure 1** *Substitution quantities of fossil (F) and RE (R) and corresponding carbon tax levels modeled using CES function for a constant output and variable: initial shares (α), elasticities of substitution (σ), and rates of decrease of RE cost (ζ).*

The absolute factor input values, i.e. F-R curve, is virtually never revealed in practice in the IAM results. Nevertheless, it is an important and tangible indicator that can be used to examine how aligned model forecasts are with historical experience. In this setting, it contradicts any reasonably expected technical substitution behavior. In our examples, when the global average F share is 85%, in order to reduce it down to 55%, it requires RE sources to be increased 13 times (for $\sigma=1.5$). This occurs progressively as for every unit of fossils that is substituted, it requires more and more RE to replace the same amount of fossils due to the decreasing marginal rate of technical substitution of RE with respect to them. We highlight the fact that this result is contrived as it is estimated with absolutely no reference to the technical capabilities and specifications of the two resources as we further discuss in Section 5.

By transforming physical inputs to prices in Eq. 2, economists and modelers manage to move away from the physicality of the technical areas to the more abstract realm of prices. It is much harder to argue with prices although their exponentially increasing behavior is as much a baseless artifact of the CES application. The carbon tax trajectory shows that it becomes exponentially harder to switch from fossils to renewables. This is not even a 100% substitution but barely a 50% one, yet carbon taxes reach 1000 times the unit cost of the fossil energy resource. The exponential rise in the price of F needed in order to force its replacement by RE corresponds to the exponential increase in the quantities of RE "needed" to replace a unit of F. Essentially, the use of prices veils the CES-driven behavior of the physical quantities. Of course, no assumption of increases in the prices of fossil resources would compensate for such radical increase in the tax levels even considering depletion or other market effects.

Since projected future energy demand has an increasing total output, we also include a model with a constant growth rate in the energy demand in Fig. 2. The effects discussed above are now exaggerated on both the quantities and the prices side as the fossil output needs to keep increasing marginalizing the effect of the RE and making it that much more difficult to reach a high substitution level.

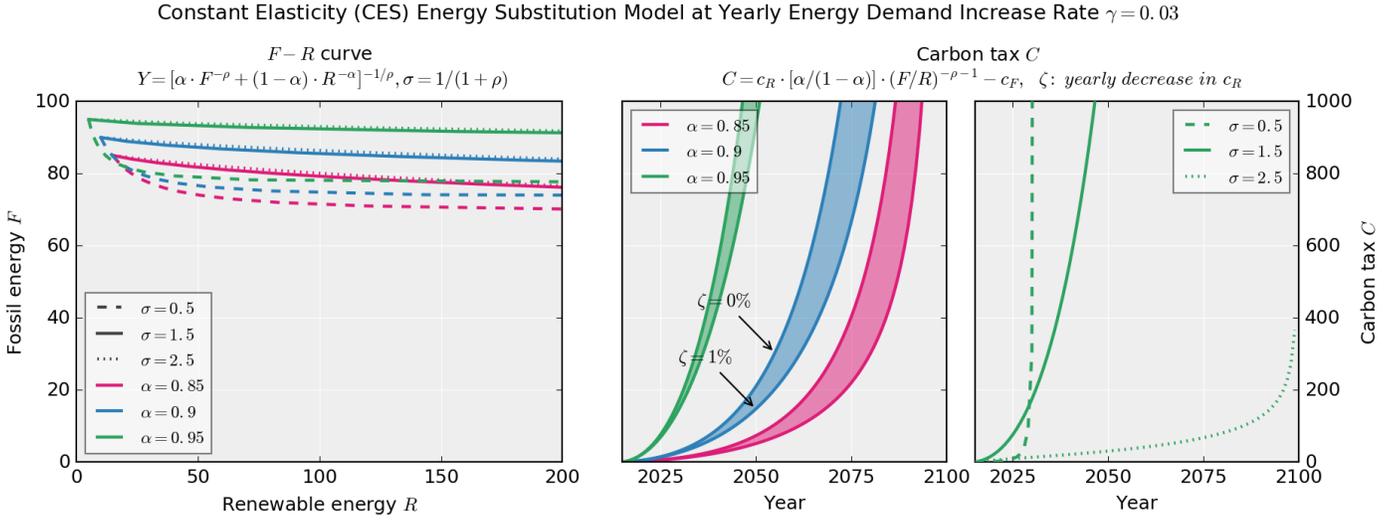

*Figure 2 Substitution quantities of fossil (F) and RE (R) and corresponding carbon tax levels modeled using CES function for increasing output at rate (γ)=3% and variable: initial shares (α), elasticities of substitution (σ), and rates of decrease of RE cost (ζ).*

This stylized example of a CES application demonstrates the fundamental characteristics and sensitivity of CES and highlights the fact that the modeled behaviors, e.g. the exponentially increasing carbon price, are artifacts not based on the characteristics and dynamics of technological substitutions. Yet these artifacts are featured as primary results in several IAMs as discussed next.

## 4 Examples of the CES Technology Transition Fallacy and its implications

(Cai et al. 2015) developed a hybrid CGE model to study global carbon tax and the pathways to mitigate carbon emissions. Their state-of-the-art, hybrid model, called GTEM-C, was designed to combine the top-down (CGE) model with the bottom-up engineering details of energy production and consumption based on a previous the version of Global Trade and Environment Model (GTEM). The improved version uses a technology bundle approach in addition to the classical CES. For electricity generation, the bundle includes nuclear, hydro, wind, solar, biomass, waste, and other renewables in addition to fossil fuels. GTEM-C uses different variants of the CES function in several calculation stages, including switching among factor productions for the total economic output but also among different technologies of energy generation. As shown in Figure 3(a), GTEM-C utilizes a combination of Leontief functions, constant return of elasticities homothetic (CRESH) functions – a CES variant (Hanoch 1971), and typical CES to create a nested structure for industrial output. The entire edifice is structured using CES.

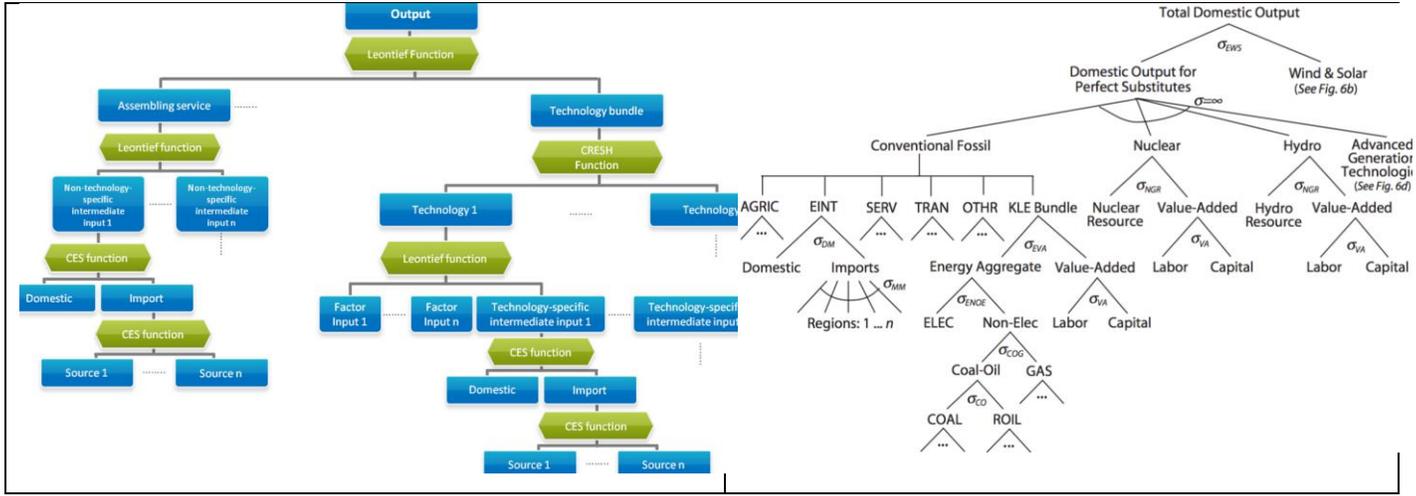

*Figure 3 Nested CES structures for (a) Industrial Production in GTECM-C utilizing technology bundles (Source: (Cai et al. 2015)) and (b) energy sector (Source: (Paltsev et al. 2005))*

It is unsurprising, therefore, that the results behave similarly to our example in Section 3. They find the need for exponentially increasing carbon taxes reaching as high as 700-900 USD/ton CO2eq in order to induce 15-20% penetration of RE in the final energy consumption although the electricity system comes close to phasing out fossil fuels shown in Figure 4a. A prominent carbon price knee towards the end of the century is the result of the hybrid approach where a different function takes over in the form of backstop (see related line in Figure 4).

Another well-regarded model that exhibits the same artifact behavior is the MIT Emissions Predictions and Policy Analysis (EPPA) v4 (Paltsev et al. 2005). EPPA is a multi-regional CGE model that intends to incorporate global economy interactions, trade and GHGs abatement dynamics using a carbon or general tax as the main policy lever. It forms the economic backbone for IGSM, a broader IAM. The model has been updated so that it includes "*a wider range of advanced energy supply technologies, improved capability to represent a variety of different and more realistic climate policies, and enhanced treatment of physical stocks and flows of energy".* Nevertheless, it is constructed using the CES function and its variants (Leontief and Cobb-Douglas) for all type of production output, including energy, and final consumption. Figure 3(b) shows the characteristic nested structure of CES function and variants used for the regional economic output (a) and energy sector (b).

As can be intuited from its structure, EPPA results follow the expected carbon price pattern. They rise exponentially to very high levels (5500 $/tCO2 or about seven times higher than in previous paper as shown in Fig. 4) in order to bring carbon emissions back to 1990 levels by 2100 and allegedly stabilize CO2 concentration at 550 ppm. The treatment of RE in EPPA is "*as imperfect substitutes represent[ing] the unique aspects of these renewable technologies. While they can be well-suited to some remote locations, they also suffer from intermittency that can add to their cost if they were to provide a large share of electricity production.*" The elasticity of substitution chosen permits gradual but limited adoption only as prices of other technologies rises. Recognizing a limitation in the treatment they refer to work by (Cheng 2005) on integrating intermittent sources like wind in CGE modeling. The approach proposed by Cheng may have merit but the entirely counterintuitive result from its application is edifying. Cheng's assumption that all intermittent sources require fossil back-up, i.e. ignoring non-fossil storage options, leads to the elimination of all such sources in a high carbon price scenario by the end of the century relying solely on nuclear, hydro and carbon-capture fossil options!

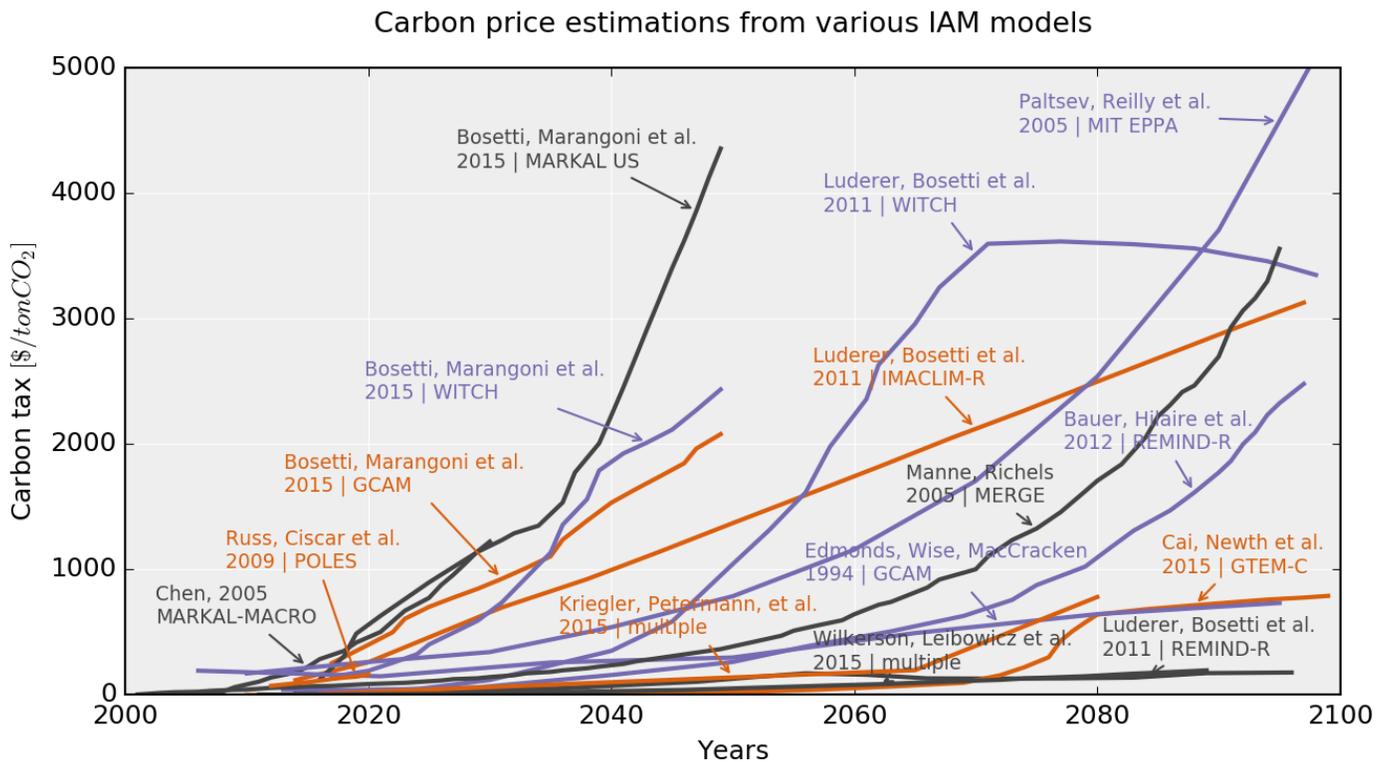

**Figure 4** *Carbon price estimations from various IAM models. For the carbon price curves presented here we assume the following conditions (unless specifically mentioned otherwise here or in the source publication. If there are multiple estimates, we use the RCP2.6 scenario or the*

*lowest ppm concentration reported. If ranges or confidence intervals are reported, the mean is taken. If the model reports price in dollar per ton of carbon, it is converted to dollar per ton of carbon dioxide.*

An example with similar results utilizes again a hybrid framework to model the dynamics of a country's GHGs emission dynamics. It combines a social welfare utility model (Macro) with a bottom-up approach for energy production (Markal) for an application on China (Chen 2005) but with the CES function used for the aggregate production (Eq. 4). The MAC cost curve for China shows comparatively moderate carbon taxes but for moderate level of abatement where growth of nuclear energy is restrained (cf. Fig. 4).

Recently efforts are under way to compare the sensitivity across IAMs in a consistent way. They do find differences but not substantial ones and generally do not question the fundamental, technical validity of the results. In these studies the carbon price becomes exogenously set and the resulting behavior of the technical inputs is monitored. (Wilkerson et al. 2015) contrasted three major models (GCAM, MERGE and EPPA) and their results are similar. Using either a quadratic increase in carbon tax or a carbon shock all three models show that the primary energy input in the future remain fossil fuels albeit with CCS. This study is notable in that it shows extreme rate changes of technical solutions seemingly without recognizing the physical impossibility of the task; GCAM shows a quadrupling of nuclear energy within 5 years, while MERGE deals with the shock by increasing biomass with CCS ten-fold within a decade. Along similar lines, (Kriegler et al. 2015), compared a larger number of models but the relevant results remain similar. Using the metric of transformation index (with 0 no change and 2 total change) the final energy mix scores values from 0.3 to 1.1 and primary energy mix from 0.9 to 1.6 with GCAM scoring the highest value with carbon prices reaching to more than 1700$/tonne. While the detailed technology mix is not clearly disclosed in this setting, by comparing the GCAM projections of the Wilkerson et al. paper we can assume that the bulk of this "transformation" is based on the growth of nuclear and biomass with CCS. The fact that biomass is considered by IAMs as much more costly to replace than RE is reinforced in a recent model survey (Luderer et al. 2013).

The IAMs we have discussed so far are very consistent in finding the need to exponentially increase the carbon tax reaching several thousands constant dollars per tonne towards the end of the century. The policy interpretation of these fallacious results can come from the modelers themselves. Early on, Nordhaus found that essentially doing nothing is an optimal policy and then proceeding to expand on entirely unsubstantiated claims of the low cost option of climate engineering (Nordhaus 1992). Since then researchers may refrain from making explicit recommendations letting their carbon price findings to speak for themselves. We identify three main problems resulting from relying on these IAM results for energy policy: *price diversion, sticker shock, and technological myopia.*

Price diversion refers to the tendency to leave all physical planning aside in favor of price setting. Since all economic models dealing with technological projections divert the discussion towards the question of pricing as a policy mechanism it is natural for the policy makers to adopt this thinking. While the social cost of carbon may be an interesting if highly indeterminate number (cf. (Allen and Frame 2007) and the discussion in Section 2), using it to guide policy decisions in a very dynamic technological transition environment where the new technology costs drop faster than any technology forecast and the price of the commodities fluctuate randomly is counterproductive. This is so because it creates the false impression that setting a price (perhaps even a large increase in the future c.f. (Wilkerson et al. 2015)) would be sufficient to meet the targets. When what is needed is physical system planning relying on setting energy availability targets and phase-out strategies that inform the business sector, instill trust in the transition reality, and fuel reinforcing economies of learning and scale.

Sticker shock is the reaction to the exponentially increasing requirement of carbon prices and the estimated "costs" of mitigations. Policy making stakeholders instinctively perceive that raising carbon taxes to the thousands when it is barely possible to pass legislation for such a rate in the low teens is politically futile. When humans are faced with what seems an impossible task subconsciously take refuge to resignation. Sisifus' optimal strategy is to stop pushing the rock up, same as for average students graded on a curve. Resignation though is the least helpful psychological attitude when dealing with what is a civilizational crisis. Additionally, a high carbon price is directly related to a perception of a high cost of alternatives. This prevents policy stakeholders from undertaking long-term technology-based planning.

Finally, this attitude reinforces the effects of technology myopia. In IAMs that improperly use the CES function, the costs of incorporating new technologies (i.e. adding a non share-preserving possibility) becomes prohibitive and they revert to the technology shares that already exist: fossil with CCS, nuclear, or biomass. This becomes facilitated not because nuclear, CCS, or biomass are viable technologies or competitive or able to scale in the necessary time frame but simply because such technologies are already included in the original formulations of the production mix and it is easy to model them with a known increase in cost. This dynamic is responsible for the ubiquitous but spurious finding that CCS including biomass with CCS (i.e. negative emissions) is a significant part of any viable technology mix for achieving a low concentration pathway, c.f. Section 4.1 but also the RCP2.6 scenario in (van Vuuren et al. 2011). When policy makers review the results they are likely to inherit the models' technology myopia failing to recognize the potential of targeted policy to influence the cost expectations.

## 5    Alternative Approaches and Indicative Estimates of Physical Substitution Costs for SET across Sectors

Energy and economy support each other in a reinforcing relationship (Ayres and Warr 2005) therefore in order to start addressing this flawed approach in technological transitions, a review of empirical findings should be a critical first step. Nevertheless, (Rosen and Guenther 2015) found "*no literature comparing investment decisions for energy-consuming equipment implicit in IAMs with real-world trends in the past.*" This may be because if the economics of IAMs were correct, when applied at past transitions, the costs of substituting the last steam locomotive or whale-oil lamp would have been tremendous but, of course, they were not. In fact the more substitution progressed, the transition accelerated becoming easier and faster until when close to saturation it started slowing down again. The slow down was not price induced, i.e. the incumbent technology was not becoming cheaper in response, but a reflection of dynamics of niche markets and social inertia. This S-curve behavior of technological transitions, portrayed in its simple form by the logistic function, is well established across economic sectors (Schumpeter 1943; Christensen 1997) but also in energy system transitions   as shown in Figure 5 (Ayres et al. 2003), (Fouquet 2010), (Grubler et al. 1999). In the widest documented sampling of technology deployment trajectories, the pattern of exponential deployment rate and exponential cost reduction in technology diffusion was consistently observed (Nagy et al. 2013).

The recognition of such dynamics is not foreign to economists – in fact it was explored early on using the logistic function and its variants (e.g. (Mansfield 1961)). The logistic function has been used successfully to describe with excellent fits historical energy transitions (Marchetti and Nakićenović 1979) on a technological/social basis but also on a physical depletion basis with the well-known Hubbert curve (Hubbert 1956; Maggio and Cacciola 2009). Other approaches include: the Gompertz-curve, a generalized version of logistic function with asymmetric early diffusion and saturation slopes commonly used from biology (Kozusko and Bajzer 2003) to technological forecasting (Martino 2003), and differential equation-based models. The most well-known such model is the Bass Diffusion model (Norton and Bass 1987), originally developed for single product adoption in a fixed-population market – an analogue of which would be the energy system and competing technologies. The solutions of the Bass equation contain both the logistic function and the Gompertz-curve as special cases.

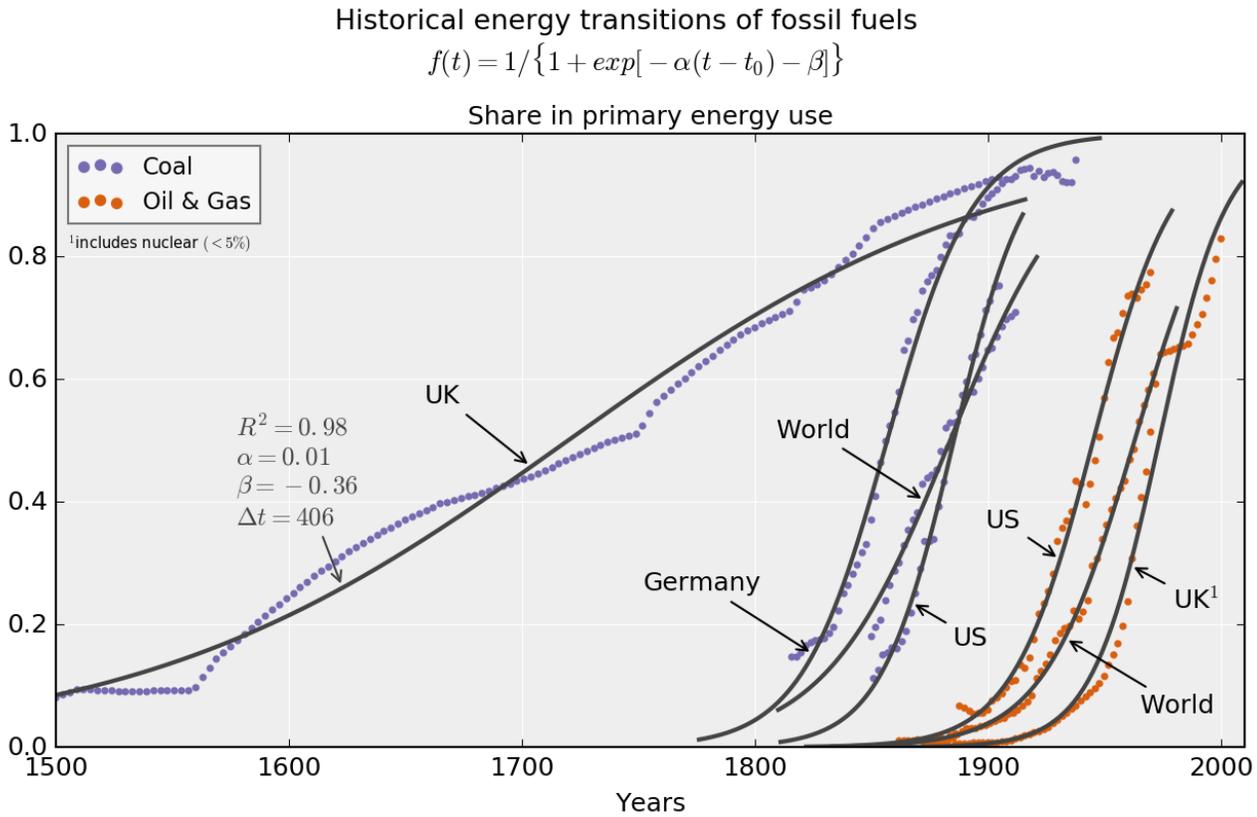

**Figure 5 Historical Energy Transitions Data and S-curve fits (Data sources: (Fouquet 2010), (Grubler 2012), (Marcotullio and Schulz 2007)))**

Unfortunately, an S-curve transition cannot be modeled with a CES function as it fundamentally presents a relationship with a *dynamically* evolving elasticity. Taking this as a starting point, we find that it is possible to better approach S-curve transition dynamics in the existing modeling formulations by dynamically adjusting the elasticity of substitution. Using the sample formulation in Section 3.2, we demonstrate this effect using both linearly and exponentially decreasing elasticities (see Figure 7 for the constant energy demand case and Fig. 8 for the increasing demand). We believe that this approach calibrated against the expected evolution of the technical profiles of the substitutes offers a viable compromise that can maintain the relevance of models that modify their CES formulations.

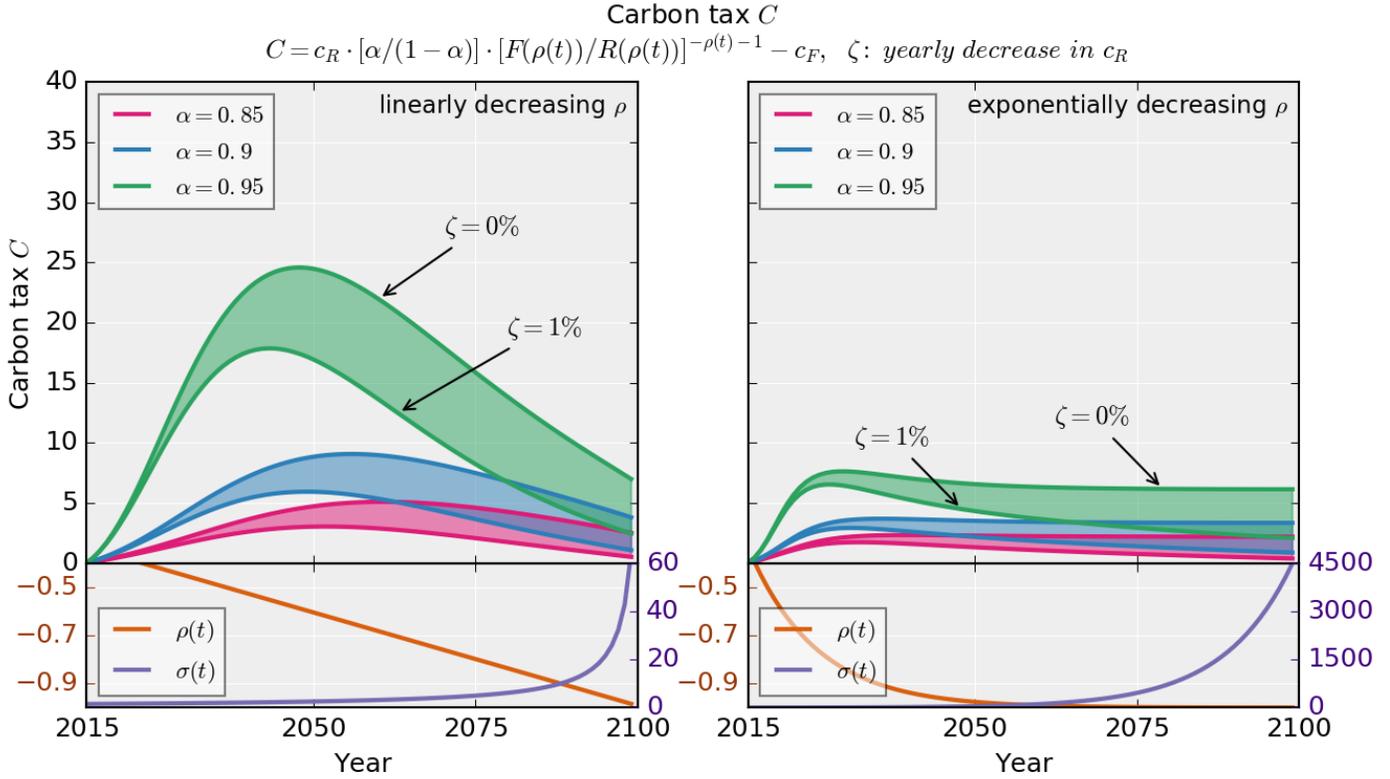

**Figure 6 Using a Dynamic Elasticity Energy Model with linear and exponential decrease of the elasticity (ρ) for constant output**

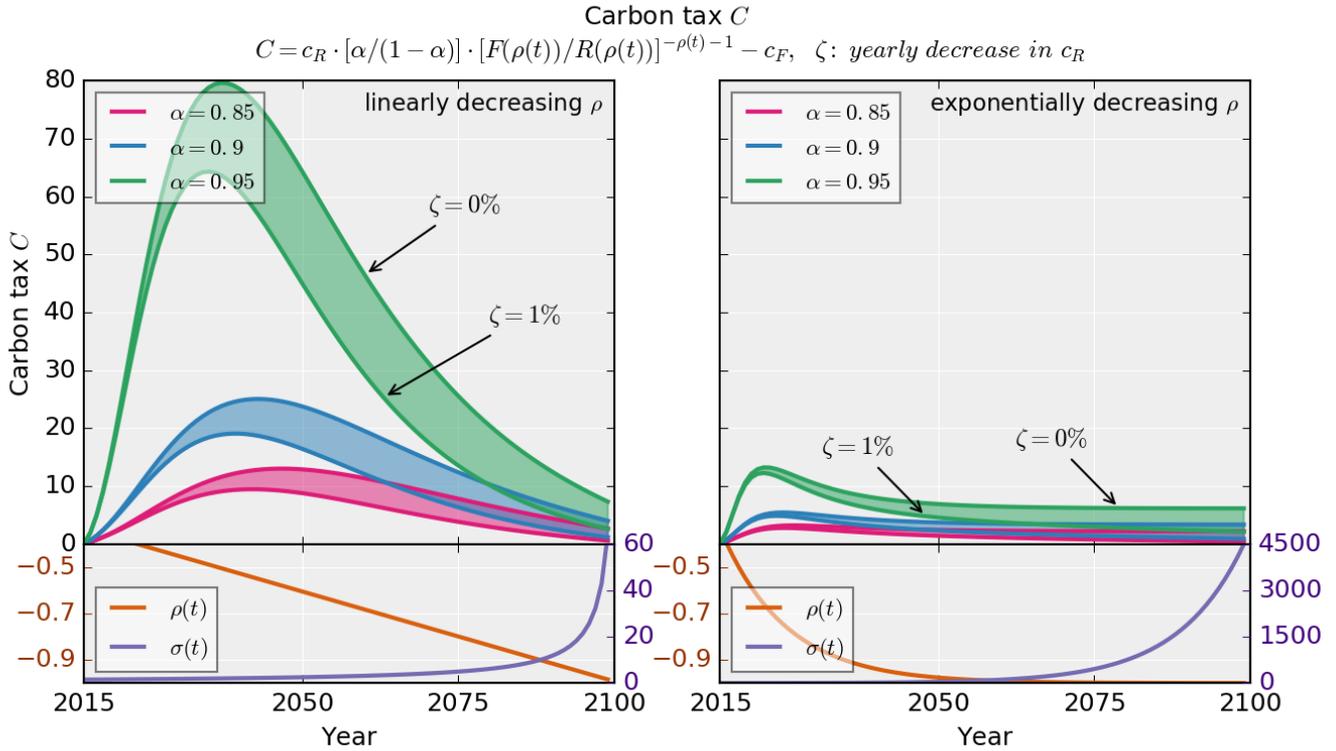

**Figure 7** *Using a Dynamic Elasticity Energy Model with linear and exponential decrease of the elasticity (ρ) for increasing output*

While a carbon price peak may be counterintuitive to economists' marginalist thinking, in practice, it is what should be expected based on the dynamics of technical diffusion. Economies of scale, scope, learning, and the initial adopters pull of followers form parts of the technology diffusion process. At the absolute worst case, the carbon price would be capped by the cost of the substitute technology even if it *is* inferior and therefore would have no reason to become exponential. Even today the back-stop alternative technologies have either reached, or are within reach of direct competition with fossil fuels on several fronts while maintaining a strong momentum towards cost reductions.

Change is inherent in an energy system that is undergoing a transition. Efficiency and demand shifting are already happening and opportunities to reduce waste are actively sought. In key sectors like transportation, changes like a modal shift from air travel to high-speed rail provide important avenues to utilize the renewable energy supply (Sgouridis et al. 2010) competing directly and even entirely subsuming the air service once the rail connections become operational for distances less than 500km. For automobiles and residential energy storage battery technologies are well on their way to become directly competitive (Nykvist and Nilsson 2015). On the supply side, the levelized cost of energy (LCOE) trajectories for wind and even more for solar photovoltaic have been consistently declining becoming directly competitive with wholesale electricity prices in several regions. While it is correctly pointed out that LCOE of RE does not account for the full cost of renewables integration (Joskow 2011), it is insufficient to rely on the estimated costs from ramping up the fossil-fuel back-ups as cost indicators. This approach takes as a given assumption that storage is expensive but is missing the point that for a SET system, the combined LCOE of RE plus short and long-term storage including power to liquids are fully representing the system operation costs and there are no additional "integration" costs. Studies that take this physical reality into consideration confirm the existence of manageable upper bound at a global (Pleßmann et al. 2014), or country level analyses e.g. Germany (Ueckerdt et al. 2015) and Japan (Komiyama and Fujii 2015). In the former study that disadvantaged renewables by not considering long-distance transmission options and the use of hydro resources, the costs of 100% RE-based do not exceed double current retail cost of electricity for the majority of the world population. For the critical long-term, large-scale storage where the proven solutions like pumped-hydro is not possible, the available options include grid-batteries, and power to liquids. These would use electrolysis to hydrogen and then hydrogen to liquids processes. The latter provide avenues for creating drop-in alternatives for applications that require energy dense fuels in addition to acting as grid support.

Expecting the carbon price to grow by several factors higher than this back-stop cost has no physical basis. Of course, as these technologies diffuse their cost would go down. Even if they do so at slower rate than the RE supply technologies have exhibited recently, their downward sloping costs imply the same for any carbon price. In this section we presented a modification to dynamic elasticity of substitution models that could be a useful way for bridging the development of physics-driven, high-resolution energy models.

# 6    Conclusions

IAMs require a paradigm shift in order to begin providing policy-relevant results that reflect the practical implications of finite resource pools in economic systems that are accustomed to exponential growth. A first critical step would be recognize the problems in the current structures and stop replicating them under the excuse of waiting for a better alternative. In practical terms, our ability to model future economic dynamics and its interactions are very limited if we rely on this "growth mentality" as it inherently prevents us from seeing pivot points. The use of CES functions to model physio-technical transitions that are by definition following a process with dynamic elasticity of substitution is a paradigmatic manifestation of such a mentality, as it reinforces the fiction that changing the current energy mix becomes exponentially harder as we proceed. This can only be associated with vertiginous increases in the carbon price. An alternative approach, should recognize that policy making is beyond simply setting a high price. By this we do not imply that the energy transition can occur unassisted in the timeframe dictated by risk-averse climate change mitigation. On the contrary, recognizing that there are institutional, political, and infrastructural barriers, governance and policy making should act by setting appropriate deployment target of RE and storage alternatives thus accelerating the dynamics of energy transitions historically observed.

### Supplementary Material:

### Marginal Abatement of Fossil Energy under a Carbon Tax by Using CES Function

Derivation of Relative Price of Fossil Energy (F) with Respect to Renewable Energy (R) and Price Elasticity of Substitution

Amount of Fossil Energy in use (Unit Energy): F

Price of Fossil Energy (\$/E): $P_F$

Amount of Renewable Energy in use (Unit Energy): R

Price of Renewable Energy (\$/E): $P_R$

Two-good (F and R) and perfectly competitive market setting: F*$P_F$ + $P_R$*R = I

Following Arrow et al. (1961):

Energy Production Function (CES): $Y = [\alpha * F^{-\rho} + (1 - \alpha) * R^{-\rho}]^{-\frac{1}{\rho}}$ (1)

Elasticity of Substitution: $\boldsymbol{\sigma} = \frac{1}{1+\rho}$ (2)

Elasticity of substitution shows how it is easy to substitute among two sources (F, R).

If:

1. $\rho$ approaches to negative infinity ($-\infty$), then two sources become perfectly complementarity (Leontief Function)
2. $\rho$ approaches to positive infinity ($+\infty$), then two sources become perfectly substitute to each other (Linear Function) and hence very easy to substitute
3. $\rho$ approaches to zero (0), then two sources become Cobb-Douglas substitutes

Lagrangian Function: $\mathcal{L} = \left[\alpha * F^{-\rho} + (1-\alpha) * R^{-\rho}\right]^{-\frac{1}{\rho}} + \lambda * \left[I - (F * P_F + P_R * R)\right]$  (3)

First-order conditions of optimal amount (F, R) without any restriction on the income imply that:

$$\frac{\partial \mathcal{L}}{\partial F} = \frac{1}{-\rho}\left[\alpha * F^{-\rho} + (1-\alpha) * R^{-\rho}\right]^{-\frac{1}{\rho}-1} * (-\rho) * \alpha * F^{-\rho-1} - \lambda * P_F = 0$$

$$\frac{\partial \mathcal{L}}{\partial R} = \frac{1}{-\rho}\left[\alpha * F^{-\rho} + (1-\alpha) * R^{-\rho}\right]^{-\frac{1}{\rho}-1} * (-\rho) * \alpha * R^{-\rho-1} - \lambda * P_R = 0$$

$$\lambda = \frac{\left[\alpha*F^{-\rho}+(1-\alpha)*R^{-\rho}\right]^{-\frac{1}{\rho}-1}*\alpha*F^{-\rho-1}}{P_F} = \frac{\left[\alpha*F^{-\rho}+(1-\alpha)*R^{-\rho}\right]^{-\frac{1}{\rho}-1}*\alpha*R^{-\rho-1}}{P_R}$$

Which implies that:

$$\frac{P_F}{P_R} = \frac{\alpha*F^{-\rho-1}}{(1-\alpha)*R^{-\rho-1}} = \frac{\alpha}{1-\alpha}*\left(\frac{F}{R}\right)^{-\rho-1}$$  (4)

Rearranging Eq. 4 to solve for the F/R ratio this equation yields:

$$\frac{F}{R} = \left(\frac{(1-\alpha)}{\alpha}*\frac{P_F}{P_R}\right)^{\frac{-1}{1+\rho}}$$  (5)

The price elasticity of substitution of F with respect R can be found as:

$$\epsilon_{F,R} = \frac{\frac{\partial F/R}{F/R}}{\frac{\partial P_F/P_R}{P_F/P_R}} = \frac{P_F/P_R}{F/R} * \frac{\partial\left(\frac{F}{R}\right)}{\partial\left(\frac{P_F}{P_R}\right)} = \frac{\frac{P_F}{P_R}}{\frac{F}{R}} * \frac{\partial\left(\left(\frac{(1-\alpha)}{\alpha}*\frac{P_F}{P_R}\right)^{\frac{-1}{1+\rho}}\right)}{\partial\left(\frac{P_F}{P_R}\right)} = \frac{\frac{\alpha}{1-\alpha}*\left(\frac{F}{R}\right)^{-\rho-1}}{\frac{F}{R}} * \left(\frac{(1-\alpha)}{\alpha}\right)^{\frac{-1}{1+\rho}} * \frac{\partial\left(\frac{P_F}{P_R}^{\frac{-1}{1+\rho}}\right)}{\partial\left(\frac{P_F}{P_R}\right)}$$

$$= \frac{\alpha}{1-\alpha}*\left(\frac{F}{R}\right)^{-\rho-2}*\left(\frac{(1-\alpha)}{\alpha}\right)^{\frac{-1}{1+\rho}}*\frac{-1}{1+\rho}*\left(\frac{P_F}{P_R}\right)^{\frac{-1}{1+\rho}-1} = \frac{-1}{1+\rho}*\left(\frac{(1-\alpha)}{\alpha}\right)^{\frac{-1}{1+\rho}-1}*\left(\frac{(1-\alpha)}{\alpha}*\frac{P_F}{P_R}\right)^{\frac{2+\rho}{1+\rho}}*\left(\frac{P_F}{P_R}\right)^{\frac{-1}{1+\rho}-1} = \frac{-1}{1+\rho}*$$

$$\left(\frac{(1-\alpha)}{\alpha}\right)^{\frac{-1}{1+\rho}-1+\frac{2+\rho}{1+\rho}}*\left(\frac{P_F}{P_R}\right)^{\frac{2+\rho}{1+\rho}}*\left(\frac{P_F}{P_R}\right)^{\frac{-1}{1+\rho}-1} = \frac{-1}{1+\rho}*\left(\frac{(1-\alpha)}{\alpha}\right)^{\frac{-2-\rho}{1+\rho}+\frac{2+\rho}{1+\rho}}*\left(\frac{P_F}{P_R}\right)^{\frac{2+\rho}{1+\rho}+\frac{2+\rho}{1+\rho}}$$

$$\epsilon_{F,R} = \frac{\frac{\partial F/R}{F/R}}{\frac{\partial P_F/P_R}{P_F/P_R}} = \frac{-1}{1+\rho}$$  (6)

$\epsilon_{F,R} = -\frac{1}{1+\rho}$, namely a 1 % increase in price of F relative to the price of R, would decrease the demand of F by $\frac{1}{1+\rho}$ relative to the R.

Carbon Tax

We find carbon tax for abatement of fossil energy by:

1. First decreasing gradually and exogenously amount of fossil fuel (F) for energy production in (1)
2. Calculating amount of R needed for decrease in F by using equation (1)
3. Finding relative price of F ($P_F$) with respect to the price of R ($P_R$) based on the quantities available for each technology by using equation 4. We assume three different cost scenarios for the $P_R$ as no decrease, or 0.5 % decrease per year, 1 % decrease per year. We also assume that cost of F is fixed throughout the lifetime of its use.

$$P_F = P_R * \frac{\alpha}{1-\alpha} * \left(\frac{F}{R}\right)^{-\rho-1}$$

4. Carbon tax at period t will be equal to the price differential of relative price of F at period t ($P_{F,t}$) to the relative cost of F at period t ($C_{F,t}$ which is assumed to be fixed and labelled as $\overline{C_F}$):

$$C_t = P_{F,t} - C_{F,t} = P_{F,t} - \overline{C_F}$$

Sectoral transition illustration

### Historical energy transitions of fossil fuels in the United Kingdom

$$f(t) = 1 / \left\{ 1 + exp[-\alpha(t-t_0) - \beta] \right\}$$

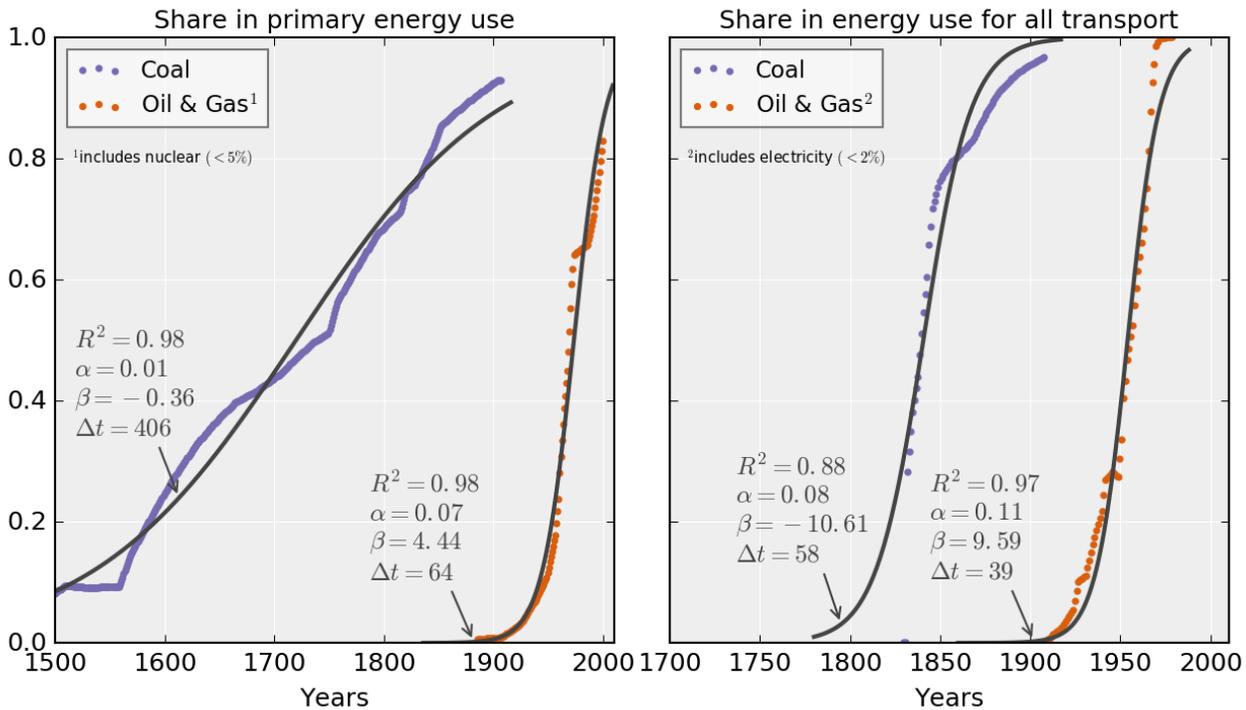

Higher order transition curve example, leading to better fit

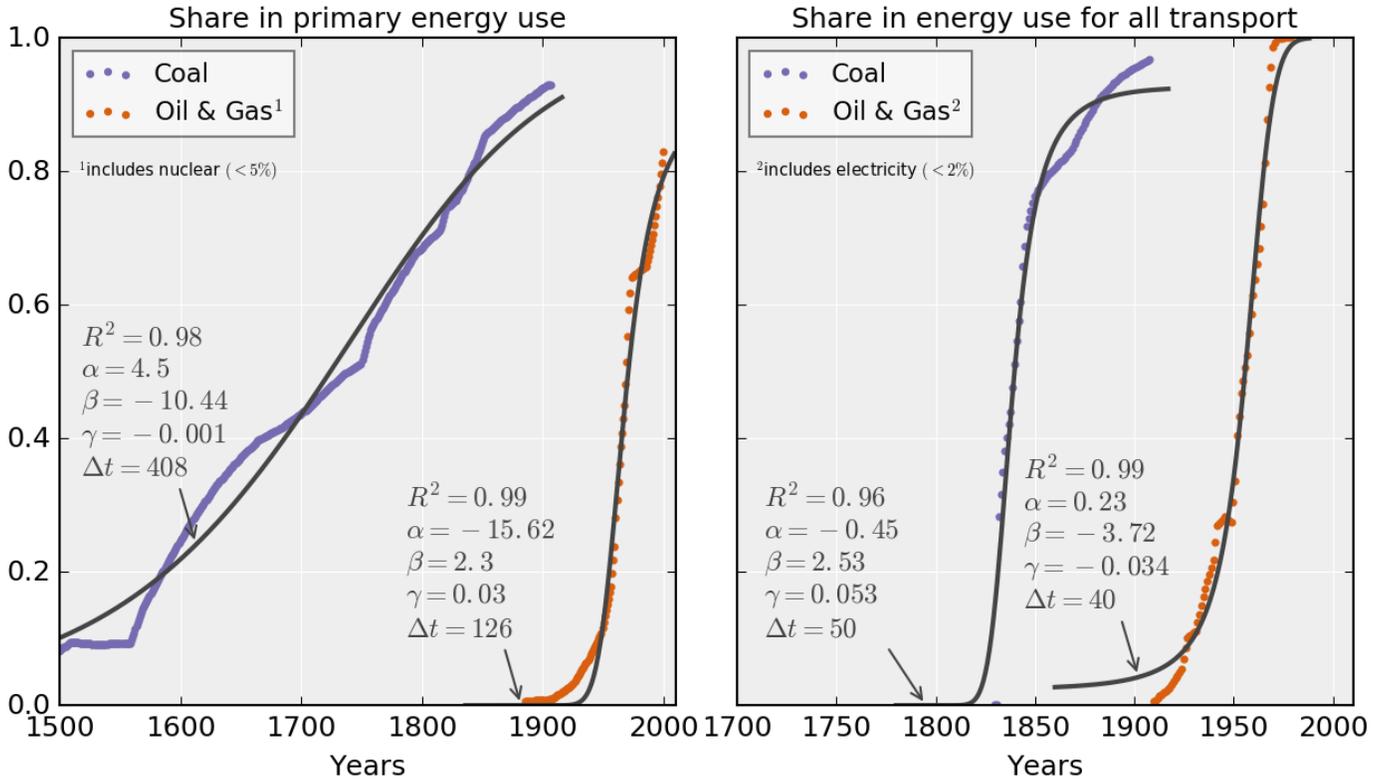

### Historical energy transitions of fossil fuels in the United Kingdom

$$f(t) = 1/\left\{1 + exp[-\alpha \cdot exp(-\gamma(t-t_0)) - \beta]\right\}$$